# Why Early-Stage Software Startups Fail: A Behavioral Framework

Carmine Giardino, Xiaofeng Wang, and Pekka Abrahamsson

Free University of Bolzano, piazza Domenicani 3 39100 Bolzano, Italia

**Abstract.** Software startups are newly created companies with little operating history and oriented towards producing cutting-edge products. As their time and resources are extremely scarce, and one failed project can put them out of business, startups need effective practices to face with those unique challenges. However, only few scientific studies attempt to address characteristics of failure, especially during the early-stage. With this study we aim to raise our understanding of the failure of early-stage software startup companies. This state-of-practice investigation was performed using a literature review followed by a multiple-case study approach. The results present how inconsistency between managerial strategies and execution can lead to failure by means of a behavioral framework. Despite strategies reveal the first need to understand the problem/solution fit, actual executions prioritize the development of the product to launch on the market as quickly as possible to verify product/market fit, neglecting the necessary learning process.

## 1 Introduction

Software startups launch worldwide every day as a result of an increase of new markets, accessible technologies, and venture capital [1]. With the term *software startups* we refer to those temporary organizations focused on the creation of high-tech and innovative products[1], with little or no operating history, aiming to grow by aggressively scaling their business in highly scalable markets [2].

New ventures such as *Facebook, Linkedin, Spotify, Pinterest, Instagram, Groupon* and *Dropbox*, to name a few, are examples of startups that evolved into successful businesses. Despite many success stories, many software startups fail before they have fulfilled their commercial potential [3].

Even though startups share some characteristics with similar contexts (e.g. small and web companies), the combination of different factors makes the specific development context quite singular [2,4].

However, failures of startups received little attention [5]. Despite the quick proliferation of startups' communities, they have been able to absorb little more than the basic patterns of how to build a startup [4]. Moreover, more than 90% of startups fail, due primarily to self-destruction rather than competition [6].

---
[1] With the term "product", we refer to both software products and software services.

This study aims to understand the software startups' dimensions, which impact failed software startups. The failure has been determined from the point of view of their chief executive officers (CEOs), who have broad perspectives on their startup organization [7].

This study was elaborated based on a multiple-case design, implemented with in-depth narratives of the two project cases, covering a wide spectrum of themes and iteratively adjusting the direction of the research according to the emerging evidence.

The results show that the two startups didn't follow consistent strategies to understand the problem they were trying to solve, consequently diluting their focus on running in the wrong direction. Despite conventional dimensions emerged to be important to improve, such as *Market, Team, Product, Business*, strategies and executions of their development were not consistent to the state of their problem/solution fit, presented by means of a behavioral framework. The behavioral framework provides a potential reason for the failure of software startups as the result of the analysis of the extrapolated data.

The rest of this paper is composed as follows: in section 2, background and the related work are presented according to the relevant disciplines. Section 3 presents the empirical research design which is followed by the presentation of the case studies results in section 4. The conclusions are identified and discussed in section 5. The paper is summarized with section 6 addressing the limitations and identifying future research needs.

## 2 Background and Related Work

In this section knowledge on the definition of software startups and how they differ from established companies is presented according to existing literature. Subsection 2.1 examines characteristics of startup companies. Subsection 2.2 describes how the state-of-the-art presents constraints, which impact the company's survival. Subsection 2.3 presents dimensions to consider when building a startup company.

### 2.1 Singularity of Software Startups

Startup companies exhibit many characteristics which reflect both engineering and business concerns. Constraints of those characteristics differ from those of established companies [5].

Established companies present advantages with respect to starutps, such as fewer internal communication and coordination problems, a foundation of established products, partners, and customers with a greater shared history and vision [4]. Sutton [4] provides a characterization of software startups, defined by the challenges they face with:

- little or no operating history: startups have little accumulated experience in development processes and organization management.
- limited resources: startups typically focus on getting the product out, promoting the product and building up strategic alliances.
- multiple influences: pressure from investors, customers, partners and competitors impact the decision-making in a company. Although individually important, overall they might not converge to support a clear decision-making.
- dynamic technologies and markets: the newness of software companies often requires them to develop or operate with disruptive technologies[2] to enter into a high-potential target market.

## 2.2 Failure Assessment

Modern entrepreneurship, born more than thirty years ago [9], has been boosted by the advent of consumer Internet markets in the middle of the nineties and culminated with the notorious dot-com bubble burst of 2000 [10]. Today, with the omnipresence of the Internet and mobile devices, we are witnessing to an impressive proliferation of software ventures- metaphorically referred to as the startup bubble. Easy access to potential markets and low cost of service distribution are appealing conditions for modern entrepreneurs [11]. Inspired by success stories, a large number of software businesses are created everyday. However, the great majority of these companies fail within two years from their creation [3].

Despite the believe that the success rate of startups has the potential to dramatically increase economic growth on global scale [6], very few and controversial findings about their failures have been found by the researchers in the last years [5].

A prominent contributor as researcher and practitioner of the startup community is Steve Blank. In his research [2] he describes how very few startups fail for lack of technology, rather they almost always fail for lack of customers. For a company trying to enter a very innovative market without proof of functionality in the real world there are more chances of failure. Customer feedback is assumed to reduce the perceived risk in a software startup. Especially in the marketing of complex and software-intensive products, experimental knowledge is crucial. However the use of customer feedback by software startup companies is distinctly under-researched [11]. Startup companies have reported that they use the first customer feedback to develop the product further, find arguments for sales and marketing purposes, learn project skills, and study the business logic in their industry [12,13]. The focus of success moves to the abilities of finding the first customers and expanding the business abroad, after saturating the limited size of local markets.

In the course of attracting and keeping customers, Blank suggests a process to place aside to product development, which aims to discover and validate the right market for an idea. The first part of the discovery consists of finding the right problem/solution fit. The aim is to test the riskiest hypotheses of the problem taken in consideration by implementing a first solution. The second step is to build the right product features that solve real customers' needs, also known as the product/market fit. If the product/market fit is not achieved, then a problem/solution fit must be reiterated, operation known as *pivoting*. Ultimately, in order to grow, companies have to test their business models, investing capital and executing tactics for acquiring and converting more customers of their potential market.

---

[2] A new technology that unexpectedly displaces an established technology. It does not rely on incremental improvements to an already established technology, but rather tackles radical technical change and innovation [8].

### 2.3 Dimensions for the Evaluation of Software Startups

Subsection 2.2 examined the customer development methodology to describe the objectives to scale a business concept. When resources are scarce, survival and success depend most heavily on the executives and managers, who are responsible for shaping, directing, and implementing company strategies. The importance of people in these roles derives from the need to keep the company focused and moving ahead [4]. However the customer development methodology doesn't define the focus in which "ahead" lies. Despite the direction might shift continually due to the dynamic and unpredictable context in which startups operate [5], some dimensions at all stages of the life cycle remain crucial.

Draw upon the study of MacMillan et al. [14], applied in startup contexts, four holistic dimensions are taken in consideration. The team as the core element. Software project managers have long recognized the importance of good people for successful development [5]. New software companies are often established to develop a technologically innovative product [4]. Yet, the business and the way it evolves can set the company growth and its place in the market [15]. The uncertainty of markets inevitably brings financial risks, which on the other hand are fundamental to set-up any company. Ultimately knowing the market is essential to evaluate the needs of the final customers [2].

## 3 Research Approach

The goal of this study aims to understand the software startups' dimensions which affect failure in a company. It is achieved through investigating how project goals are defined and decision-making is executed in the period of time that goes from idea conception to the validation of a product on the market. The following section presents the research design and rationale for methodology selection. Subsection 3.1 presents the context of the studied software startups and the short description of their business ideas.

Following a systematic mapping study (SMS) [16], we conducted an exploratory multiple-case study [17]. We executed two semi-structured interviews (with the CEOs of two failed startup companies) integrated with narrative descriptions of their failures. Moreover data were triangulated with external documentation, collected and elaborated by the founding teams. From all the data, we extracted and analyzed a model explaining the failure phenomenon of these two startups.

A first preliminary SMS in the software engineering (SE) databases revealed a quite wide gap in studies addressing failure in software startups, but at the same time it showed us how broad the domain is. A definition which confirms the suitability of this methodology is provided by Kitchenham et al. in [18]: SMSs ''are designed to provide a wide overview of a research area, to establish if research evidence exists on a topic and provide an indication of the quantity of the evidence [...]''.

Next a multiple-case study approach was chosen in view of its ability to be flexible in deepening the understanding of a specific problem, covering a wide spectrum of topics and analyzing recurring patterns among them. Going from the beginning extensively into a specific problem in software startups context would not be worth trying because we didn't find a sufficient number of studies that constituted a solid body of knowledge, confirmed by [5]. On the other hand, trying to select a broad research topic would carry the risk of spending a lot of time in achieving a better understating of the problem without being able to provide any significant results. A multiple-case design allows behavior analysis over control analysis, with the benefit of understanding the phenomena in an unmodified setting [17]. It also allows to analyze patterns across many cases and to understand how they are affected by local conditions to develop more sophisticated descriptions and powerful explanations [19].

Moreover, as we wanted the ability to transfer our ideas to startup practitioners, we applied a comparative analysis of the state-of-the-art and the case study findings following evidence-based software engineering principles [20,21] and providing empirical evidences supporting startups' decisions. The sampling of the two startups has been selected according to the features identified in subsection 2.1. Yet they are two failed evidences, as required by our research goal, as perceived from the two CEOs from their idea conception to the first open beta release.

We captured the most relevant aspects of the companies activities, letting emerge patterns from the data and adjusting the framework as we proceeded. The findings are derived primarily from the observations and perceptions of the two CEOs. The reason for selecting CEOs as respondents is that they have experienced all the decision-making process of the companies. Furthermore, they participated in the creation of the companies since their inception, having strong insights into the working activities they conducted and financed. However, the findings are finally compared to the existing literature on software startups' failure to improve generalizability.

In this paper we address the achievements of four dimensions defined by MacMillan et al. [14] (i.e. product, team, business, market).

The whole procedure was executed by the first author. Subsequently reviewed by the second and third authors. When necessary all the authors performed an in-depth review of the design of the study and discussed the findings to improve their validity.

### 3.1 Background to the Cases

The first startup (called Milkplease[3]) aims to deliver grocery shoppings from local supermarkets to the door-step of final consumers through the collaboration of neighbors. It provides a web application allowing people to know that a neighbor needed to do his shopping.

Milkplease was founded by an engineer and a computer scientist by using their private savings. Running the business for one year, they increased the team size in the last months with three more people, specialized in marketing and service design. The business model expected profits by applying a service cost, in the form of percentage over the total shopping amount for each accomplished delivery. Initially the startup targeted full-time working employees for making orders, and students for making deliveries in a small district of an European city.

The second startup (called Picteye[4]) aims to create an on-line service to sell pictures of public and private events. They provided a web market-place where people could sell and buy pictures. They started with providing pictures of one big event in Italy. Then, they moved to use the application for other big events around Europe.

The overall team grew from 3 software engineers (founders) to 6, covering business and marketing positions. Picteye has been running for two years, trying to launch the product several times without any positive response from the market. The business model expected profits by selling pictures (digitally delivered to private individuals). The startup targeted photographers for uploading pictures, and people participating to events for making orders.

## 4 Results

By analyzing the interview transcripts and narrative descriptions with the multiple-case study process, we extracted the information to describe the thematic areas of the startups and construct the final framework, presented in subsection 4.1. The thematic areas are divided according to MacMillan et al. dimensions.

This section contains quotes from the CEOs, referred with the capital "C". The trailing number indicates the software startup (e.g. "C1").

The **product** dimension characterizes the software development strategies in the two case studies. Since the two startups wanted to launch the product on the market as quickly as possible, they built an initial prototype and iteratively refined it over time, similarly to the concept of "evolutionary prototyping" [22]. However no initial hypotheses were defined from the business perspectives, and consequently no minimal viable product (MVP), as meant by Ries [23], was achieved.

"We were so excited about the technology that we put it upfront. We wanted the product to be ready for being used by many customers. The day of the launch arrived, but no one made a transaction. We still don't know why". [C1]

---

[3] Milkplease website is available at www.milkplease.it
[4] Picteye website is available at www.picteye.com

"I don't know how and why people buy photographs on-line. My startup has not answered to this question either. We were in doing mode. Ultimately 1 photo has been sold, and this was after presenting the platform in a conference, when someone in the audience saw himself in the picture gallery and decided to give the system a try". [C2]

Building a product without pursuing the problem/solution fit hasn't provided any learning process during the first period of the startup creation. In this regards, the learning based approach has been neglected by the two CEOs to benefit their attitude to acquire more and more customers. Yet, testing the initial hypotheses means have better understanding of the product risk, building the right solution for a problem worth solving.

The team dimension represents the characteristics of the people involved in building the company. The founding teams of the two startups were initially formed by full-stack engineers, able to tackle different problems at different levels of the technology stack, and covering multi-roles since they all participated in the business and development processes. However entrepreneurial characteristics were missing, such as motivation and ability to evaluate and react to risks.

"We started the project with a clear vision of what we wanted to achieve. However, we soon shifted the vision towards money concerns, refining our business model without having familiarity with the market". [C1] "During the first event week, we were presenting the project for the first time and part of the founding team bailed out. The motivation was missing". [C2]

Motivation has been felt as crucial to pursue the success of the first product release. Indeed, everyone should have been involved to understand customers' expectations and allow the entrepreneurs to evaluate and react to risks perceived during this period according to the collected feedback.

The market dimension represents the strategies for customers' acquisition. The customers of the two startups were initially acquired to pursue the product/market fit by using extensive press coverage.

"We participated to different startups' contests, where we got quickly hype. In the beginning we were sponsored by press, indeed many potential customers started visiting our web-page. We wanted all of them starting using the product from day one". [C1] "I organized an event, sending an email to all the people I could. I said few motivational words about the product. However, only few people went to the event, and shadows started growing around the project". [C2]

Customer acquisition started to be the first focus, even though the problem/solution fit was not discovered yet. With the benefit of hindsights, the CEOs argued that having small amount of participants in the beginning of the process would have allowed the startup to build an effective path to customers who care about the promoted solution, and ultimately found a market that would have supported a viable business.

The business dimension focuses on strategies to make profits out of the product. From the business perspective the two startups started to focus on maximizing the profit aspect too early, over-planning the model and prematurely adapting the market according to the business, and not vice-versa.

"Money became a first concern. As soon as we realized the product, we wanted to cover all the expenses, and maximize profits". [C1] "After the event, we started waiting the money flow in the form of completed order forms. We had everything ready for the great success". [C2]

The two cases reveal how the several dimensions were executed without regular customers' feedback loop, without adapting the business and the product to a changing market. To understand how the CEOs did not execute the strategies to capture the problem/solution fit, we make use of a behavioral framework (see figure 1).

### 4.1 The Behavioral Framework

Through analyzing the results, the two startups moved through similar thresholds and milestones of development, which we segmented into stages. From the narratives we wanted to preserve the temporal precision of the events moving along the prominent objectives. We took advantage of hindsights to see what the strategies would have been during the first period of time and what was actually executed.

The final framework (see figure 1) contains the evolution of four dimensions to build the startup, according to their exploration and validation stages. Exploration stage describes the activities focused on solving a meaningful problem, finding a viable solution. The validation stage allows to understand if the product is of interest to the potential market and if customers are willing to buy it.

The actual stage describes what they should have focused on, according to their initial strategies. The behavioral stage describes the execution of different dimensions, inconsistent to what they had initially planned. The time-line suggests how startups would have worked through the different dimensions (further analyzed in section 5).

From the product perspective a first solution should contain the minimal functional features to address the riskiest hypotheses of the business idea. Within an evolutionary approach, user experience should be improved to let customers smoothly use the product. Eventually new functionalities can be added to scale in the major market.

"We wanted to test the product on the market as soon as possible. However we had no clear hypotheses to test. We wanted to be ready for the public launch, even though we didn't evaluate known risks." [C1] "We thought and hoped that the market launch date was coming shortly to finally see the results. We knew all our thoughts were just speculations. However, I had not been thinking about lean startup method, we were entusiastic about the technology". [C2]

From the team perspective, up to when the problem and solution don't fit, the team should have entrepreneurial characteristics to be able to share the vision and evaluate the risks of the market. Only then team should start covering missing roles, such as marketing and business specialized positions.

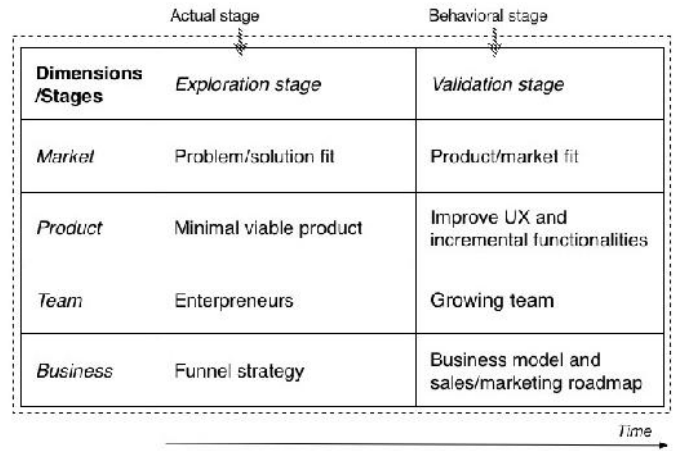

Fig. 1. Behavioral framework

"The vision was not clear as soon as we were ready to test the product. Seemingly the time spent on the product changed our expectations. The team size grew with marketing specialists and service designers". [C1] "No one was fully committed to the project. The team's motivation went down soon. However we started to grow with more specialists". [C2]

From the business perspective at the beginning the two startups should have planned how to converge traffic to satisfy and retain a small group of users. However they focused on getting the business model refined.

"The business model has been polished to maximize profits even before the product was ready. We invested our own money, and wanted them back as soon as possible. When we realized there was no market for what we realized, it was surprisingly too late to change the product again" [C1] "The company was running and it was time to get money. We asked for 20k euro to further develop the idea. We lost every month 100 euro, affordable for the first months. Soon this was not true anymore" [C2]

At the beginning the two startups tried to be profitable, rather than moving along the problem/solution fit and starting covering the need of a funnel strategy[5]. The first objective of getting profits has never been reached. Ultimately no more resources were available for possible pivots. Polishing the product and business model has consumed more resources than moving through a build-measure-learn cycle [23].

The difficulty of covering the negative cash flow, before the product or service brings revenue from real customers, has become critical over time. Up to the first launch no feedback has been given by potential customers, without effectively capturing their needs.

---

[5] A strategy that illustrates the customer journey towards the purchase of a product or service, improving the awareness of the existence of a product, activating interest and desire to choose the product [2].

"We ran the system within a group of potential users to get opinions. They helped us only with the UX". [C1] "An amateur photographer would have given us excellent feedback. He did a brief experiment, and sent me an email where he killed the system. I thought he was not the type of customer using the system. I continued waiting for the big launch." [C2]

## 5  Discussion

This section compares the overall findings of this study to existing literature, in order to emphasize crucial factors of failure in early-stage software startups. Subsection 5.1 provides insights on the lack of problem/solution fit, whilst subsection 5.2 explores the revealed weak learning process. Ultimately subsection 5.3 gives an understanding of how the framework can be adopted by practitioners and researchers.

The two described startups didn't consistently follow the defined stages. Their behavioral stage differed from their actual stage working on some operations prematurely.

As shown in the results instead of targeting the problem/solution fit, they were pursuing the product/market fit, from the *product, team, market and business* perspectives. Aiming at the product/market fit, startups focus on validating a product instead of discovering and testing a problem space. Moreover, where startups should be testing demand for a functional product, they focused on streamlining the product and making their customer acquisition process more and more efficient.

### 5.1  Lack of Problem/Solution Fit

Trying to validate a product for a problem you haven't encountered yet is a waste. Instead, being able to demonstrate problem/solution fit through discovering what are the needs of your first customers is much more viable than an untested story [24].

The two startups were improving the products' user experience without having tested the MVP first. Since 1994, Carmel [25] reports the need of flexible and rapid development solutions to shorten time-to-market to test assumptions. Eric Ries [23] shows an evolutionary approach to focus on implementing a limited number of suitable functionalities (MVP). The MVP does not require to comply with heavy quality constraints, enabling the team to quickly implement suitable functionalities to test business assumptions, ready to be validated by final users. The evolutionary approach enhances the effectiveness of the product, and enables the company's capabilities to adjust the trajectory of product development [24].

The two case companies showed initial evidences of a lack of systematic feedback from customers to improve their market understanding. However, startups face uncertain conditions, which should lead to a fast learning from trial and error, with a strong customer relationship. Avoiding wasting time in building unneeded functionalities helps also in preventing exhaustion of resources [26,27,4].

As team dimension is concerned, in order to have the focus and the ability to evaluate risks, Carmel suggests that entrepreneurs need to look for a well formed, skilled core development team, rather than just a set of product ideas and features. Moreover the team must be empowered with full-stack and self-organization settings, paired with the characteristics of real entrepreneurs. Studies suggest that entrepreneurs possess 'special' personality characteristics [28,9]. One of the key determinants of success in startup companies is the passionate behavior of the founders. People who lack passion often use the first barrier they encounter as an excuse for failure. People who have high passion will do whatever it takes to overcome those barriers. "What we can achieve in life depends on a number of things: how hard we work, how smart we work, how much leverage we have on the work we do, and how much courage we have in pursuing our goals. How hard we work is tied to how passionate we are" [29].

In addition to the missing personal motivation, the two startups were not familiar with the targeted markets. Despite a market can be uncertain, and not clear in all its aspects, a good entrepreneur can anticipate and is proactive to unforeseen events [30,31]. Instead of starting gradually with a funnel strategy of how to know the market dynamics and get the first paying customer, the startups wrongly focused on perfecting the business model.

### 5.2 Neglected Learning Process

As the firms mature and the awareness of the competitive environment grows, there is an increasing reluctance to share ideas, problems or solutions in the wider sense [32]. Indeed, over time learning progress slowed down and the two startups were only concerning how to gather more and more customers. However, improving customer acquisition before problem/solution fit has been premature, because users were not hooked to the product. Involving the customer to activate the learning progress has also been discussed by Yogendra [33] as an important factor to encourage an early alignment of business concerns to technology strategies.

The two startups progressed through their life-cycle inconsistently (see figure 1), addressing specific challenges in the wrong time, such as the need to acquire more and more customers with a growing team, maximizing profits without learning the market dynamics. These strategies prompted further challenges in trying to strike the need of investments, indeed wasting time in further structure and formalization of the business model.

Ultimately, a lost of confidence to achieve independence after using up personal savings, heavily impacted the survival of the companies, as experienced also by McAdam et al. [32]. Learning mechanisms (e.g. learning about one's strengths, weaknesses, skills attitudes etc.) have been widely researched by Cope [34], who reveals a deeper conceptualization of the process of learning from venture failures.

### 5.3 Implications of the Behavioral Framework

Existing frameworks have recognized the need of aligning business intentions to a set of development activities to achieve a certain goal (e.g. GQM Strategies [35], Balanced Scorecard [36]). However those approaches have been conceived for more established companies, taking in consideration traditional development approaches oriented towards process improvement initiatives.

The described behavioral framework, constructed by means of studies on startup cases, can be used as a tool by practitioners to analyze how consistently they are distributing their focus on the Macmillan et al. dimensions along with the exploration and validation stages. For example, a configuration might appear as figure 2, where the marked strategies (the market and team dimensions) represent a premature scaling towards the validation stage.

As remarked by Blank [2], the product/market fit has the objective of validating people's interest in a proposed product. If it doesn't occur, "pivoting" needs to be applied through changing the product or tackling a problem from a different angle. He continues reporting that if the validation phase is successful, the startup company can start increasing profits and improving their customer acquisition process (also known as scaling phase). However knowing the mismatch of possible dimensions, according to a targeted phase, can prevent the waste of resources and improve the decision process.

In real settings, this framework would allow CEOs to focus and understand how decision-making strategies are aligned to the activities within a software startup company in the early-stage of its development. Moreover, venture capitalists might better understand the stage of development of a startup company and filter those who can simultaneously sustain a consistent focus on different dimensions.

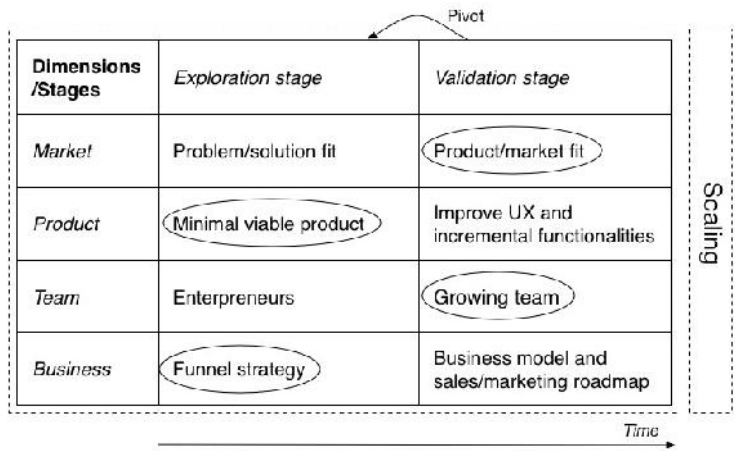

Fig. 2. Behavioral framework: the dimensional mismatch

## 6 Conclusions and Recommendation for Future Research

Software startups are able to produce cutting-edge software products with a wide impact on the market, significantly contributing to the global economy. Software development, especially in the early-stages, is at the core of the company's daily activities. Despite their severely high failure-rate, the quick proliferation of software startups is not supported by a scientific body of knowledge [5]. This paper provides an initial explanation of failure by means of a multiple-case studies based on two software startups, focusing on early-stage activities, from the *market*, *product*, *team* and *business* perspectives.

The behavioral framework is derived from the hindsight knowledge collected from the CEOs of the two failed startups, with the aim of explaining how inconsistent decision-making strategies could lead to failure.

One important validity threat to this study is the small number of cases. However, as described by Klein et al. [37], there is a basis for abstraction and generalization in interpretive field studies through the use of ideas and concepts if rigorously collected and experienced by researchers. As suggested by [38], to validate the explanatory capability and correctness of the model we compared the findings with the state-of-the-art (see section 5). To validate interview data we examined also supporting evidences to verify their expressed opinions, such as emails, presentations and documentation. The two studied startups might also be biased by contextual factors, such as type of product, competitive landscape etc. To mitigate this threat we constructed the framework using Macmillan et al. dimensions, widely used in previous software engineering studies [30,24], enabling a broader reasoning related to the factors that hinder the success of software startups.

The overall results of our study reveal inconsistency between the strategy of understanding and testing the problem/solution fit and the behavioral execution of pursuing the product/market fit. When resources are scarce, survival and success depend most heavily on the executives and managers, who are responsible for shaping, directing, and implementing company strategies. Early recognition and management of critical issues can increase the chances of success for a software startup. The two startups failed to understand the problem and provide the right solution, showing increasing reluctance in learning from the potential customers.

In this paper we integrated a number of novel challenges discovered according the Macmillan et al. dimensions [14] for both practitioners and researchers, while presenting a first set of concepts, terms and activities which set startup strategies for the rapidly increasing startup phenomenon. By means of the behavioral framework, we provided a possible reason for the failure of software startups. Consequently, there is a great deal of scope here for further research. For example, further research is required to investigate the following two outstanding questions; Firstly, how to prevent mismatch between business intentions and development execution? Secondly, how existing learning processes can improve business and development alignment?